\documentclass[aps,prl,reprint,groupedaddress,longbibliography]{revtex4-1}

\usepackage{amssymb,amsfonts,amsmath}
\usepackage{graphicx } 
\usepackage{dcolumn}   
\usepackage{bm}        
\usepackage{color}     

\begin{document}
\newcommand{\comment}[1]{}
\newcommand{\Orf}{\Omega_\textrm{rf}}
\newcommand{\Sr}{$^{88}$Sr$^+$ }
\newcommand{\kB}{\textrm{k}_\textrm{B}}
\newcommand{\ket}[1]{\left| #1 \right>} 
\newcommand{\bra}[1]{\left< #1 \right|} 
\newcommand{\braket}[1]{\left< #1 \right>} 
\newcommand{\p}[2]{\left|#1\left>\right<#2\right|} 
\newcommand{\level}[3]{\textrm{#1}_{#2}\left(#3\right)}

\title{Direct observation of atom-ion non-equilibrium sympathetic cooling}

\author{Ziv Meir}
\email[Current affiliation: Department of Chemistry, University of Basel, Basel 4056, Switzerland. Email:]{ziv.meir@unibas.ch}
\author{Meirav Pinkas}
\author{Tomas Sikorsky}
\author{Ruti Ben-shlomi}
\author{Nitzan Akerman}
\author{Roee Ozeri}
\affiliation{Department of Physics of Complex Systems, Weizmann Institute of Science, Rehovot 7610001, Israel}
\date{\today}

\begin{abstract}
Sympathetic cooling is the process of energy exchange between a system and a colder bath. 
We investigate this fundamental process in an atom-ion experiment where the system is composed of a single ion, trapped in a radio-frequency Paul trap, and prepared in a coherent state of $\sim$200 K and the bath is an ultracold cloud of atoms at $\mu$K temperature. 
We directly observe the sympathetic cooling dynamics with single-shot energy measurements during one, to several, collisions in two distinct regimes. 
In one, collisions predominantly cool the system with very efficient momentum transfer leading to cooling in only a few collisions.
In the other, collisions can both cool and heat the system due to the non-equilibrium dynamics of the atom-ion collisions in the presence of the ion-trap's oscillating electric fields. 
While the bulk of our observations agree well with a molecular dynamics simulation of hard-sphere (Langevin) collisions, a measurement of the scattering angle distribution reveals forward-scattering (glancing) collisions which are beyond the Langevin model.
This work paves the way for further non-equilibrium and collision dynamics studies using the well-controlled atom-ion system.
 
\end{abstract}

\maketitle
Sympathetic cooling is a key paradigm in the physics of ultracold atoms and ions. With ions, sympathetic cooling is used to cool ionic species which lack laser-cooling transitions \cite{Larson1986}, and molecules \cite{Mølhave2000,Blythe2005,Tong2010}, for the purpose of creating new time standards \cite{Chou2010}, testing fundamental physics theories \cite{Schiller2005,Rosenband2008} and performing chemistry at extremely low temperatures with a well defined quantum state and a single particle resolution \cite{Ratschbacher2012,Sikorsky2017a,Furst2017}. In the case of ultracold neutral atoms, sympathetic cooling is used to achieve degeneracy in species with small \cite{Myatt1997} or vanishing collision cross section as in the case of low temperature fermions \cite{Truscott2001}. Recently, sympathetic cooling with atoms was extended to cool a macroscopic membrane \cite{Jockel2015}. Here, we study the dynamics of sympathetic cooling of a single ion by an ultracold bath of atoms with single-collision and single-event resolution.  

Experimental systems overlapping ultracold atoms and ions are an excellent resource for studying not only sympathetic cooling \cite{Zipkes2010,Schmid2010,Rellergert2013} but also non-equilibrium dynamics \cite{BlueSky,Meir2016}. The absence of equilibria emanates from the oscillating electric fields which are used to trap the charged ions \cite{Paul1990}. These fields couple to the ion's motion during a collision \cite{Dehmelt1968} and can lead to both heating and cooling regardless of the colliding atom's energy. The non-equilibrium dynamics results in a power-law distribution of the ion energies \cite{DeVoe2009,Zipkes2011,Chen2014,Pascal2016,Rouse2017,Meir2017exp} which was recently observed experimentally using Rabi spectroscopy on a narrow linewidth transition \cite{Meir2016}. In that work, the ion's energy distribution was inferred from a steady-state Tsallis function fit to a mean signal of many experimental realizations. Here, we study the non-steady-state regime following single to few collisions and with direct ion energy measurements in each experimental realization, thus using no a-priory assumptions on the ion's energy distribution. 

We use a newly developed thermometry technique \cite{Meir2017cooling} which is similar to Doppler-cooling thermometry (DCT) \cite{Wesenberg2007,Sikorsky2017}, however, utilizes different analytic tools and is suited for a different energy regime. DCT is used to extract the temperature of the ion assuming an underlying energy distribution and requires extensive averaging over many experimental realizations due to photon shot noise. In the new method, single-shot Doppler-cooling thermometry (SSDCT), we extract the energy of the ion from each experimental realization, without averaging, leading to the direct reconstruction of its energy distribution with no underlying assumptions. The method relies on detecting the cooling time which is independently assigned for every repetition of the experiment. This method is also capable of identifying the spatial mode of ion motion in every experimental realization with good statistical significance, an ability we exploit to measure the distribution of scattering angles.

\begin{figure}
	\centering
	\includegraphics[width=\linewidth,trim={1cm 1cm 2cm 2cm},clip]{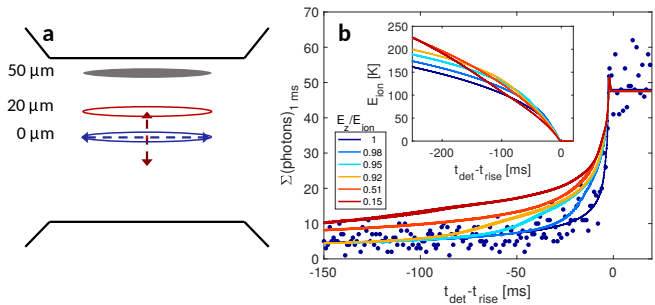}\\
	\caption{\textbf{a) Two experimental configurations.} The ellipse indicates the size and position of the atomic cloud with respect to the ion-trap center. Gray is the initial position before interaction, blue (red) is the position in the axial (radial) equilibrium (non-equilibrium) experiment. Dashed arrow lines indicate the initial coherent-motion amplitude of the ion in each of the experiments. Arrows and ellipses are to scale. \textbf{b) Single-shot Doppler-cooling thermometry.}  The dots are measured fluorescence signal of an ion prepared solely in the axial mode of the trap. The lines are results of a numerical calculation of the ion's fluorescence for different initial axial to radial energy ratios (legend - E$_\textrm{z}$/E$_\textrm{ion}$ is the axial to total energy ratio). Inset is the total ion's energy for the different numerical calculations. In all graphs, we set the zero time to the onset of reaching the Doppler temperature.}
	\label{fig:systemandsimulation}
\end{figure}

We performed two distinct experiments. In both experiments, we prepared a single ion in a classical coherent state of $\sim$200 K with distribution width of few K. In the first experiment, we prepare the coherent state along the axial mode of the ion trap and we overlap the atoms with the center of the trap. In the second experiment, we prepare the coherent state in the radial mode and we offset the atoms 20 $\mu$m above the center of the ion trap (see Fig. \ref{fig:systemandsimulation}a). In symmetric linear Paul traps there are no oscillating electric fields along the axial axis of the trap such that collisions only cool the ion. Moreover, the atomic cloud finite size in the radial direction (5 $\mu$m) introduces an energy cutoff for the distribution power law of few K \cite{Dutta2017,Pascal2016} which allows the observation of equilibrium dynamics in this inherently-non-equilibrium system. In the second experiment, the ion is excited radially and samples regions of the trap with high-amplitude oscillating fields. Collisions with atoms in the presence of these fields lead to non-equilibrium effects. Since the electric field amplitude in an ion Paul trap increases linearly with the radial distance from the trap center, we displace the atoms radially from the center of the trap to allow for collisions in regions of high amplitude fields.

Our experimental apparatus is described in details in \cite{Meir2017exp}. Shortly, we prepare an ultracold atomic sample (5 $\mu$K), consisting of $\sim$7,500 atoms, in a crossed optical dipole trap, 50 $\mu$m above the center of the ion trap (Fig. \ref{fig:systemandsimulation}a). The ion is Doppler cooled and micromotion compensated at the trap center. We then excite the ion using a short oscillating electric field pulse close to resonance with the axial (or radial) mode frequency. This results in a narrow Gaussian distribution of energies \cite{Meir2017cooling}. 
At this point, we move the atoms to overlap the ion. We scan the interaction time \cite{SM} to observe the effect of single to few collisions on the ion's energy distribution. To stop the interaction we shut off the atom's dipole trap beams which leads to a fast decrease of the atomic density and halts atom-ion collisions. We then use the SSDCT to detect the ion energy and mode of motion for the specific experimental instance. This method (SSDCT) is described in detail in \cite{Meir2017cooling} and will only be briefly discussed in the following. 

In SSDCT, we detect the ion's fluorescence during the course of Doppler cooling 
(blue dots in Fig. \ref{fig:systemandsimulation}b). At the initial stage of cooling, the ion's fluorescence is low due to the large Doppler shifts associated with the ion's coherent motion. The photon-scattering level remains low for 100's ms until the Doppler shifts become comparable with the cooling transition linewidth. At this point, a sharp rise of the fluorescence signal indicates the last stage of cooling where the ion reaches the Doppler temperature limit. From a numerical calculation of the ion's trajectory in the presence of the cooling light, we determine the ion's scattering rate dynamics and compare it to the experimental measured fluorescence. 

We repeat the numerical calculation for different energy partitions between the axial and radial modes of the ion motion. The results of these numerical calculations are shown in Fig. \ref{fig:systemandsimulation}b. We see that the initial partition of energy between the trap modes changes significantly the fluorescence dynamics. For energy distribution purely in the axial mode of the trap (dark blue line in Fig. \ref{fig:systemandsimulation}b) the fluorescence level at high energies is low compared to the case of energy distributed considerably also in the radial modes (redder lines). We also notice that the fluorescence rise to the Doppler level at the last cooling stage is much sharper for the axially distributed energy. Both effects are due to the broadening of the spectrum due to micromotion sidebands \cite{Cirac1994,Sikorsky2017}. The micromotion sidebands, which appear only in the radial modes, both increase the fluorescence at high energies and decrease the cooling rate at lower energies. The inset of Fig. \ref{fig:systemandsimulation}b shows the evolution of the ion's total energy during process of cooling. We see that the initial mode partition of energy also changes the cooling time. For ion prepared in the axial mode of the trap (dark blue lines) the cooling time is longer compared to the more radially populated cases (redder lines). This effect is attributed to the increased scattering due to micromotion sidebands discussed before and to the cooling and trap geometry. The evolution of the energy in the different modes of the trap is given in \cite{SM}.

We perform a likelihood ratio-test (LRT) for the fluorescence curve of each single-shot experimental repetition with all the calculated numerical fluorescence trajectories and choose the one with the maximal likelihood value (see \cite{SM} and \cite{Meir2017cooling}). We performed a stochastic numerical simulation to analyze the LRT method. The type I errors for assigning a wrong simulation to a trajectory are less than 1\% when the cooling trajectory is longer than 100 ms \cite{SM}. We use this numerical calculation to extract the ion's energy for the specific experimental instance and repeat the procedure for all other instances.

\begin{figure}
	\centering
	\includegraphics[width=\linewidth,trim={0cm 4cm 0cm 1cm},clip]{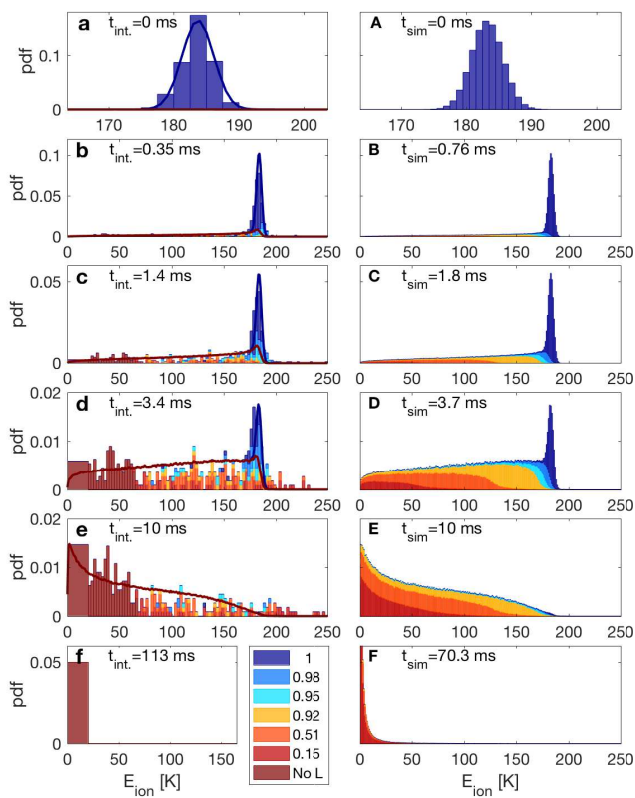}\\
	\caption{\textbf{Sympathetic cooling.} a-f) Experimentally measured energy histograms of the ion's total energy for different interaction times, t$_\textrm{int.}$ as defined in \cite{SM}. The color of the bars indicates the axial energy to total energy ratio of the ion according to the LRT analysis (see legend). The color code is the same as in Fig. \ref{fig:systemandsimulation}b. The dark red bars are cold events with cooling times less than 50 ms which the LRT cannot determine the mode distribution \cite{SM}, however, the SSDCT can determine the energy. We use the steady-state energy distribution (Ez/E$_\textrm{ion}$=0.15), calculated using MD simulation, for these events. The lines are results of MD simulation which takes into account only Langevin collisions which are also presented in the right column. In the left column, the red line separates between simulation events with no collisions (above) and with one or more collisions (below). A-F) MD simulation energy histograms. The color of the bars indicates the axial energy to total energy ratio in the simulation binned according to the available numerical simulations. Simulation times, $t_\textrm{sim}$, are chosen to match the measured mean energy in the experiment.}
	\label{fig:axialexpresults}
\end{figure}

The results of the experiment in which the ion motion is prepared along the axial mode and the atoms overlap the center of the trap are shown in Fig. \ref{fig:axialexpresults}a-f. The measured initial energy histogram of the ion before collisions with the atoms is shown in Fig. \ref{fig:axialexpresults}a. The ion is in a classical coherent state with a mean energy of 183.7 K. The Gaussian distribution width of 2.3 K is either due to imperfect initialization of the coherent state or noise in the energy measurement owing to fluctuations in the parameters of the cooling beams. The narrow distribution sets an upper limit for the SSDCT stability of $\sim$1\%. The accuracy of this method was estimated to be on the level of 5$\%$ in \cite{Meir2017cooling}. The LRT identified 98.4$\%$ of the experimental events energies to distribute solely in the axial mode (dark blue) which sets an upper bound of 1.6\% for the type II error for assigning a collision to no-collision event.

\begin{figure}
	\centering
	\includegraphics[width=\linewidth,trim={0cm 0cm 0cm 0cm},clip]{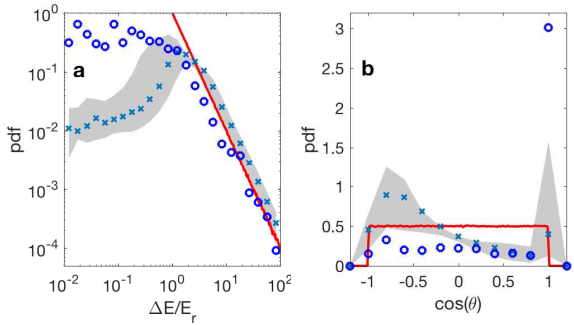}\\
	\caption{\textbf{Scattering angle distribution.} Histograms of the energy difference divided by the radial energy after a collision (a) and the cosine of the scattering angle (b). Red lines are the theoretical prediction of Langevin scattering. Blue exes are analysis of a MD simulation results (Fig. \ref{fig:axialexpresults}A-D). The analysis takes into account the experimental limitations which modify the distributions. Gray areas reflect the uncertainty region \cite{SM} of the simulation analysis due to the experimental limitations. The simulation treats only Langevin collisions. Blue circles are calculated from the experimental data of Fig. \ref{fig:axialexpresults}a-d.}
	\label{fig:cross_section}
\end{figure}

As the ion starts colliding with the atoms (Fig. \ref{fig:axialexpresults}b-e) we see that the energy histogram develops a tail of low energies and that the initial narrow Gaussian distribution depletes. This is a direct observation of sympathetic cooling in which a collision with essentially zero energy atom can take away all the energy of the ion (we use $^{87}$Rb atoms and a $^{88}$Sr$^+$ ion with nearly equal masses). We also see that the mode of the collided events is no longer solely in the axial direction but contains a substantial fraction of the energy in the radial direction as well (the bars color indicates the fraction of radial energy). This is a direct observation of the collision scattering angle. A quantitative analysis of the scattering angle distribution is given in Fig. \ref{fig:cross_section} and is discussed below. We observe a small excess of collisions which result in energies above the ion's initial energy. These seemingly non energy conserving events are probably due to over estimation of the ion's energy for large scattering angles (see Fig. \ref{fig:systemandsimulation}b inset).

In previous work \cite{Meir2016}, we measured the energy distribution in steady-state of our system to be a Tsallis power-law distribution with an energy scale in the mK regime. The SSDCT method is only sensitive to energies above 20 K. Even for the high power-law tail, $P(E)\propto E^{-2.4}$, of our system, with almost diverging mean energy, the probability to detect an event above 20 K in steady-state is negligible (3$\cdot10^{-4}$\%). Since we did not detect experimentally any of such events in 440 trials (Fig. \ref{fig:axialexpresults}f), we can put an upper bound to the probability of these energetic events at 0.23\% in agreement with our previous work.

We perform a molecular-dynamics (MD) simulation, similar to the one described in \cite{Zipkes2011}, to extract the energy and the energy mode distribution of the ion. We use the experimentally measured initial energy distribution (Fig. \ref{fig:axialexpresults}a) as the initial energy distribution in the simulation (Fig. \ref{fig:axialexpresults}A). To account for uncertainties in the density of the atoms which are on the order of 50\%, we compare the experimentally measured mean energy after different interaction times to the simulated one to adjust the simulation time, $t_\textrm{sim}$, to the experimental time, $t_\textrm{int.}$ (the experimental and simulation time are given in Fig. \ref{fig:axialexpresults}). The simulated energy histograms are shown in Fig. \ref{fig:axialexpresults}B-F together with the distinction between the different energy partitions between modes of motion (color). We add the simulation results also to the experimental data figures (Fig. \ref{fig:axialexpresults}a-f solid lines). Here, the red lines separate between events which have undergone a Langevin collision and events which did not. In Fig. \ref{fig:axialexpresults}c-d we see a small excess of experimentally detected events in which motion was detected to be at a small angle to the longitudinal direction (light blue) compared to the MD simulation predictions (red line). This observation indicates that our thermometry method is sensitive to glancing collisions which are not taken into account in the simulation. This effect is also observed in the second experiment in which motion was initially excited along the radial direction.

To study this effect further we perform a quantitative analysis of the scattering angle distribution, shown in Fig. \ref{fig:cross_section}. For the case of a free ion moving at a constant velocity, colliding with a zero-energy atom, the energy transferred to the atom, $\Delta E$, and the ion deflection angle, $\theta$, are correlated. We can neglect the ion's trapping harmonic potential in the collision analysis since the collision is instantaneous as compared with the ion harmonic trap frequency. However, due to the harmonic potential, we can only experimentally measure the total energy (kinetic and potential) of the ion at the moment the first collision occurs and not its velocity. We therefore use an observable, $\Delta E/E_r$, which is independent of the velocity of the ion prior to the collision (see \cite{SM}),
\begin{equation}\label{eq:costheta}
    \cos(\theta)=\left(\frac{m}{M}-\left|\frac{\Delta E}{E_r}-\frac{m}{M}\right|\right)/\frac{\Delta E}{E_r}.
\end{equation}
Here, $m$ ($M$) is the ion (atom) mass and $E_r$ is the radial part of the ion energy after the collision where we assume that the initial energy is solely in the axial direction. In Langevin collisions, $\cos(\theta)$, where $\theta$ is the scattering angle in the center-of-mass frame, is distributed isotropically (red lines in Fig. \ref{fig:cross_section}). However, our experimental limitations, which are mainly due to the binning of the radial energy due to the LRT mode analysis, result in a modified distribution indicated by the blue exes and shaded gray area in Fig. \ref{fig:cross_section}. The experimental results of the data in Fig. \ref{fig:axialexpresults}a-d is given in blue circles. For more details see \cite{SM}.

Fig. \ref{fig:cross_section} shows a good agreement between the experimental and the theoretical scattering angles for small angles ($0<\cos\left(\theta\right)<1$), however, it deviates from theory in the forward scattering ($\cos\left(\theta\right)\approx1$) and backward scattering ($\cos\left(\theta\right)<0$) regimes. The forward-scattering deviation is due to glancing collisions which are not part of the Langevin model. The backward scattering deviation is due to the collision large angle and small resulting ion's energy in the lab frame which are not compatible with our LRT analysis.  

In a second experiment, we initialize the ion in a coherent state along one of the radial modes of the ion trap. The ion's trajectory samples regions of the trap with increasing oscillating electric fields. These fields can couple to the ion's kinetic energy during a collision which leads to heating even when colliding with essentially zero temperature atom \cite{DeVoe2009,Zipkes2011,Chen2014,Pascal2016,Rouse2017}. This non-equilibrium dynamics is manifested by a power-law tailed energy distribution in steady state \cite{Meir2016}. Here, we focus on the single collision dynamics where the system is far from steady-state. To enhance the non-equilibrium effect, we displace the atoms 20 $\mu m$ away from the trap center (Fig. \ref{fig:systemandsimulation}a) such that collisions occur in a region of high-amplitude oscillating electric fields.

\begin{figure}
	\centering
	\includegraphics[width=\linewidth,trim={0cm 6.5cm 0cm 1cm},clip]{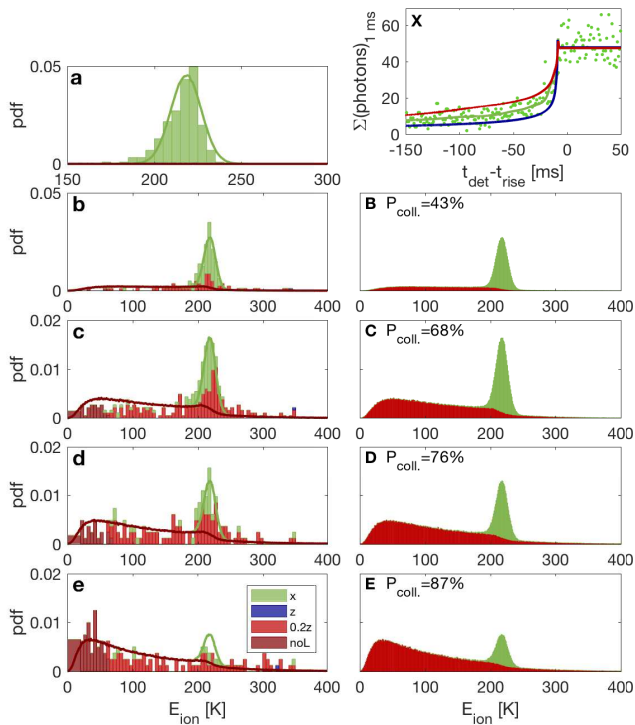}\\
	\caption{\textbf{Non-equilibrium dynamics.} X) Dots are measured single-shot fluorescence signal where the ion's energy is initialized in the x-radial mode of the trap. Lines are results of numerical calculations where the ion's energy is initialized in the x-mode (green), z-mode (blue) and the steady-state mixture (0.4/0.45/0.15 - x/y/z) of modes. a-e) Experimentally measured total energy histograms for increasing interaction time. The bars' color code indicates the result of the LRT and it is the same as in (X). Dark red bars are low-energy events which are below the LRT sensitivity. The energy value of all events is extracted using the x-mode numerical calculation. Lines are results of a MD simulation. The area underneath the red line represents events with at-least single collision while the area above the red line signals events which did not collide according to the simulation. A-E) MD simulation total energy histograms. Color code indicates events which are more than 99$\%$ x-mode (green) or z-mode (blue). The rest of the events are colored in red. We adjust the MD simulation time to the experimental time according to the percentage of no-collision events (green) the value of which is given for (B-E).}
	\label{fig:radialexpresults}
\end{figure}

The experimental results are shown in Fig. \ref{fig:radialexpresults}a-e. The ion is initialized in a classical coherent state of 218.2$\pm$8.8 K along the x-radial mode of the trap (green in Fig. \ref{fig:radialexpresults}a). As we overlap the ion and atoms, the SSDCT identifies single collision events (red in \ref{fig:radialexpresults}a-e) by the LRT method described above. In this experiment, we compare the experimental results to three numerical simulations, x-mode (green), z-mode (blue) and steady-state mode (red) distributions (Fig. \ref{fig:radialexpresults}X). We see that before interaction the LRT identified 99\% of the events as x-mode events (green). This gives a bound for the false collision detection error below 1\%. We also see that the LRT identified events which collided with the atoms to be almost exclusively in the steady-state mode with agreement with the MD simulation (Fig. \ref{fig:radialexpresults}B-E). These events (marked in red in Fig. \ref{fig:radialexpresults}b-e) can take values up to 350 K which are considerably larger than the ion's initial energy. Here, we use the x-mode trajectory for extracting the energy of all events including the z mode and the steady-state mode events. This way we make sure that we are not overestimating the energy as in the previous, axial, experiment. Thus, we show here a direct observation of the non-equilibrium dynamics in atom-ion collisions. The non-equilibrium dynamics we observe in the experiment is more pronounced than what is predicted by the MD simulation. We believe this is due to a non-Gaussian profile of the atomic cloud which leads to collisions in trap regions of even larger amplitude oscillating fields.  

As in the previous experiment, we see a peak of collision events with the same energy as the initial energy of the ion. These forward-scattering (glancing) events are not accounted for in the MD simulation since it includes only Langevin (spiraling) collisions. This result, together with the scattering angle analysis given in Fig. \ref{fig:cross_section}, emphasizes the sensitivity to collision directionality of the LRT and the ability to explore collision dynamics beyond the Langevin model.

To summarize, we used a novel thermometry technique (SSDCT) with single-event and energy-mode resolution to study the dynamics of sympathetic cooling of a hot ion immersed in an ultracold bath of neutral atoms. We demonstrated the capabilities of SSDCT to detect a single collision and the direction of ion motion following this collision. We used this capability to observe a deviation of the scattering angle distribution from the Langevin model predictions, manifested by a forward-scattering peak. We also directly observed the non-equilibrium dynamics of atom-ion collisions in a Paul trap with single-collision resolution. This work paves the way for using atom-ion systems as a test-bed for non-equilibrium experiments, exploiting the ultimate control over these systems parameters and their inherent non-equilibrium properties. 

We thank Eric Hudson for useful discussions. This work was supported by the Crown Photonics
Center, ICore-Israeli excellence center circle of light, the
Israeli Science Foundation and the European Research Council (consolidator grant 616919-Ionology).

\bibliography{BibFile}

\begin{thebibliography}{32}%
\makeatletter
\providecommand \@ifxundefined [1]{%
 \@ifx{#1\undefined}
}%
\providecommand \@ifnum [1]{%
 \ifnum #1\expandafter \@firstoftwo
 \else \expandafter \@secondoftwo
 \fi
}%
\providecommand \@ifx [1]{%
 \ifx #1\expandafter \@firstoftwo
 \else \expandafter \@secondoftwo
 \fi
}%
\providecommand \natexlab [1]{#1}%
\providecommand \enquote  [1]{``#1''}%
\providecommand \bibnamefont  [1]{#1}%
\providecommand \bibfnamefont [1]{#1}%
\providecommand \citenamefont [1]{#1}%
\providecommand \href@noop [0]{\@secondoftwo}%
\providecommand \href [0]{\begingroup \@sanitize@url \@href}%
\providecommand \@href[1]{\@@startlink{#1}\@@href}%
\providecommand \@@href[1]{\endgroup#1\@@endlink}%
\providecommand \@sanitize@url [0]{\catcode `\\12\catcode `\$12\catcode
  `\&12\catcode `\#12\catcode `\^12\catcode `\_12\catcode `\%12\relax}%
\providecommand \@@startlink[1]{}%
\providecommand \@@endlink[0]{}%
\providecommand \url  [0]{\begingroup\@sanitize@url \@url }%
\providecommand \@url [1]{\endgroup\@href {#1}{\urlprefix }}%
\providecommand \urlprefix  [0]{URL }%
\providecommand \Eprint [0]{\href }%
\providecommand \doibase [0]{http://dx.doi.org/}%
\providecommand \selectlanguage [0]{\@gobble}%
\providecommand \bibinfo  [0]{\@secondoftwo}%
\providecommand \bibfield  [0]{\@secondoftwo}%
\providecommand \translation [1]{[#1]}%
\providecommand \BibitemOpen [0]{}%
\providecommand \bibitemStop [0]{}%
\providecommand \bibitemNoStop [0]{.\EOS\space}%
\providecommand \EOS [0]{\spacefactor3000\relax}%
\providecommand \BibitemShut  [1]{\csname bibitem#1\endcsname}%
\let\auto@bib@innerbib\@empty
\bibitem [{\citenamefont {Larson}\ \emph {et~al.}(1986)\citenamefont {Larson},
  \citenamefont {Bergquist}, \citenamefont {Bollinger}, \citenamefont {Itano},\
  and\ \citenamefont {Wineland}}]{Larson1986}%
  \BibitemOpen
  \bibfield  {author} {\bibinfo {author} {\bibfnamefont {D.~J.}\ \bibnamefont
  {Larson}}, \bibinfo {author} {\bibfnamefont {J.~C.}\ \bibnamefont
  {Bergquist}}, \bibinfo {author} {\bibfnamefont {J.~J.}\ \bibnamefont
  {Bollinger}}, \bibinfo {author} {\bibfnamefont {Wayne~M.}\ \bibnamefont
  {Itano}}, \ and\ \bibinfo {author} {\bibfnamefont {D.~J.}\ \bibnamefont
  {Wineland}},\ }\bibfield  {title} {\enquote {\bibinfo {title} {{Sympathetic
  cooling of trapped ions: A laser-cooled two-species nonneutral ion
  plasma}},}\ }\href {\doibase 10.1103/PhysRevLett.57.70} {\bibfield  {journal}
  {\bibinfo  {journal} {Physical Review Letters}\ }\textbf {\bibinfo {volume}
  {57}},\ \bibinfo {pages} {70--73} (\bibinfo {year} {1986})},\ \Eprint
  {http://arxiv.org/abs/arXiv:1011.1669v3} {arXiv:arXiv:1011.1669v3}
  \BibitemShut {NoStop}%
\bibitem [{\citenamefont {M{\o}lhave}\ and\ \citenamefont
  {Drewsen}(2000)}]{Mølhave2000}%
  \BibitemOpen
  \bibfield  {author} {\bibinfo {author} {\bibfnamefont {K}~\bibnamefont
  {M{\o}lhave}}\ and\ \bibinfo {author} {\bibfnamefont {M}~\bibnamefont
  {Drewsen}},\ }\bibfield  {title} {\enquote {\bibinfo {title} {{Formation of
  translationally cold MgH+ and MgD+ molecules in an ion trap}},}\ }\href
  {\doibase 10.1103/PhysRevA.62.011401} {\bibfield  {journal} {\bibinfo
  {journal} {Physical Review A}\ }\textbf {\bibinfo {volume} {62}},\ \bibinfo
  {pages} {011401} (\bibinfo {year} {2000})}\BibitemShut {NoStop}%
\bibitem [{\citenamefont {Blythe}\ \emph {et~al.}(2005)\citenamefont {Blythe},
  \citenamefont {Roth}, \citenamefont {Fr{\"{o}}hlich}, \citenamefont {Wenz},\
  and\ \citenamefont {Schiller}}]{Blythe2005}%
  \BibitemOpen
  \bibfield  {author} {\bibinfo {author} {\bibfnamefont {P.}~\bibnamefont
  {Blythe}}, \bibinfo {author} {\bibfnamefont {B.}~\bibnamefont {Roth}},
  \bibinfo {author} {\bibfnamefont {U.}~\bibnamefont {Fr{\"{o}}hlich}},
  \bibinfo {author} {\bibfnamefont {H.}~\bibnamefont {Wenz}}, \ and\ \bibinfo
  {author} {\bibfnamefont {S.}~\bibnamefont {Schiller}},\ }\bibfield  {title}
  {\enquote {\bibinfo {title} {{Production of Ultracold Trapped Molecular
  Hydrogen Ions}},}\ }\href {\doibase 10.1103/PhysRevLett.95.183002} {\bibfield
   {journal} {\bibinfo  {journal} {Physical Review Letters}\ }\textbf {\bibinfo
  {volume} {95}},\ \bibinfo {pages} {183002} (\bibinfo {year}
  {2005})}\BibitemShut {NoStop}%
\bibitem [{\citenamefont {Tong}\ \emph {et~al.}(2010)\citenamefont {Tong},
  \citenamefont {Winney},\ and\ \citenamefont {Willitsch}}]{Tong2010}%
  \BibitemOpen
  \bibfield  {author} {\bibinfo {author} {\bibfnamefont {Xin}\ \bibnamefont
  {Tong}}, \bibinfo {author} {\bibfnamefont {Alexander~H.}\ \bibnamefont
  {Winney}}, \ and\ \bibinfo {author} {\bibfnamefont {Stefan}\ \bibnamefont
  {Willitsch}},\ }\bibfield  {title} {\enquote {\bibinfo {title} {{Sympathetic
  Cooling of Molecular Ions in Selected Rotational and Vibrational States
  Produced by Threshold Photoionization}},}\ }\href {\doibase
  10.1103/PhysRevLett.105.143001} {\bibfield  {journal} {\bibinfo  {journal}
  {Physical Review Letters}\ }\textbf {\bibinfo {volume} {105}},\ \bibinfo
  {pages} {143001} (\bibinfo {year} {2010})},\ \Eprint
  {http://arxiv.org/abs/arXiv:1006.5642v1} {arXiv:arXiv:1006.5642v1}
  \BibitemShut {NoStop}%
\bibitem [{\citenamefont {Chou}\ \emph {et~al.}(2010)\citenamefont {Chou},
  \citenamefont {Hume}, \citenamefont {Koelemeij}, \citenamefont {Wineland},\
  and\ \citenamefont {Rosenband}}]{Chou2010}%
  \BibitemOpen
  \bibfield  {author} {\bibinfo {author} {\bibfnamefont {C.}~\bibnamefont
  {Chou}}, \bibinfo {author} {\bibfnamefont {D.}~\bibnamefont {Hume}}, \bibinfo
  {author} {\bibfnamefont {J.}~\bibnamefont {Koelemeij}}, \bibinfo {author}
  {\bibfnamefont {D.}~\bibnamefont {Wineland}}, \ and\ \bibinfo {author}
  {\bibfnamefont {T}~\bibnamefont {Rosenband}},\ }\bibfield  {title} {\enquote
  {\bibinfo {title} {{Frequency Comparison of Two High-Accuracy Al+ Optical
  Clocks}},}\ }\href {\doibase 10.1103/PhysRevLett.104.070802} {\bibfield
  {journal} {\bibinfo  {journal} {Physical Review Letters}\ }\textbf {\bibinfo
  {volume} {104}},\ \bibinfo {pages} {070802} (\bibinfo {year} {2010})},\
  \Eprint {http://arxiv.org/abs/arXiv:0911.4527v2} {arXiv:arXiv:0911.4527v2}
  \BibitemShut {NoStop}%
\bibitem [{\citenamefont {Schiller}\ and\ \citenamefont
  {Korobov}(2005)}]{Schiller2005}%
  \BibitemOpen
  \bibfield  {author} {\bibinfo {author} {\bibfnamefont {S.}~\bibnamefont
  {Schiller}}\ and\ \bibinfo {author} {\bibfnamefont {V.}~\bibnamefont
  {Korobov}},\ }\bibfield  {title} {\enquote {\bibinfo {title} {{Tests of time
  independence of the electron and nuclear masses with ultracold molecules}},}\
  }\href {\doibase 10.1103/PhysRevA.71.032505} {\bibfield  {journal} {\bibinfo
  {journal} {Physical Review A}\ }\textbf {\bibinfo {volume} {71}},\ \bibinfo
  {pages} {032505} (\bibinfo {year} {2005})}\BibitemShut {NoStop}%
\bibitem [{\citenamefont {Rosenband}\ \emph {et~al.}(2008)\citenamefont
  {Rosenband}, \citenamefont {Hume}, \citenamefont {Schmidt}, \citenamefont
  {Chou}, \citenamefont {Brusch}, \citenamefont {Lorini}, \citenamefont
  {Oskay}, \citenamefont {Drullinger}, \citenamefont {Fortier}, \citenamefont
  {Stalnaker}, \citenamefont {Diddams}, \citenamefont {Swann}, \citenamefont
  {Newbury}, \citenamefont {Itano}, \citenamefont {Wineland},\ and\
  \citenamefont {Bergquist}}]{Rosenband2008}%
  \BibitemOpen
  \bibfield  {author} {\bibinfo {author} {\bibfnamefont {T.}~\bibnamefont
  {Rosenband}}, \bibinfo {author} {\bibfnamefont {D.~B.}\ \bibnamefont {Hume}},
  \bibinfo {author} {\bibfnamefont {P.~O.}\ \bibnamefont {Schmidt}}, \bibinfo
  {author} {\bibfnamefont {C.~W.}\ \bibnamefont {Chou}}, \bibinfo {author}
  {\bibfnamefont {A.}~\bibnamefont {Brusch}}, \bibinfo {author} {\bibfnamefont
  {L.}~\bibnamefont {Lorini}}, \bibinfo {author} {\bibfnamefont {W.~H.}\
  \bibnamefont {Oskay}}, \bibinfo {author} {\bibfnamefont {R.~E.}\ \bibnamefont
  {Drullinger}}, \bibinfo {author} {\bibfnamefont {T.~M.}\ \bibnamefont
  {Fortier}}, \bibinfo {author} {\bibfnamefont {J.~E.}\ \bibnamefont
  {Stalnaker}}, \bibinfo {author} {\bibfnamefont {S.~A.}\ \bibnamefont
  {Diddams}}, \bibinfo {author} {\bibfnamefont {W.~C.}\ \bibnamefont {Swann}},
  \bibinfo {author} {\bibfnamefont {N.~R.}\ \bibnamefont {Newbury}}, \bibinfo
  {author} {\bibfnamefont {W.~M.}\ \bibnamefont {Itano}}, \bibinfo {author}
  {\bibfnamefont {D.~J.}\ \bibnamefont {Wineland}}, \ and\ \bibinfo {author}
  {\bibfnamefont {J.~C.}\ \bibnamefont {Bergquist}},\ }\bibfield  {title}
  {\enquote {\bibinfo {title} {{Frequency Ratio of Al+ and Hg+ Single-Ion
  Optical Clocks; Metrology at the 17th Decimal Place}},}\ }\href {\doibase
  10.1126/science.1154622} {\bibfield  {journal} {\bibinfo  {journal}
  {Science}\ }\textbf {\bibinfo {volume} {319}},\ \bibinfo {pages} {1808--1812}
  (\bibinfo {year} {2008})}\BibitemShut {NoStop}%
\bibitem [{\citenamefont {Ratschbacher}\ \emph {et~al.}(2012)\citenamefont
  {Ratschbacher}, \citenamefont {Zipkes}, \citenamefont {Sias},\ and\
  \citenamefont {K{\"{o}}hl}}]{Ratschbacher2012}%
  \BibitemOpen
  \bibfield  {author} {\bibinfo {author} {\bibfnamefont {Lothar}\ \bibnamefont
  {Ratschbacher}}, \bibinfo {author} {\bibfnamefont {Christoph}\ \bibnamefont
  {Zipkes}}, \bibinfo {author} {\bibfnamefont {Carlo}\ \bibnamefont {Sias}}, \
  and\ \bibinfo {author} {\bibfnamefont {Michael}\ \bibnamefont {K{\"{o}}hl}},\
  }\bibfield  {title} {\enquote {\bibinfo {title} {{Controlling chemical
  reactions of a single particle}},}\ }\href {\doibase 10.1038/nphys2373}
  {\bibfield  {journal} {\bibinfo  {journal} {Nature Physics}\ }\textbf
  {\bibinfo {volume} {8}},\ \bibinfo {pages} {649--652} (\bibinfo {year}
  {2012})}\BibitemShut {NoStop}%
\bibitem [{\citenamefont {Sikorsky}\ \emph
  {et~al.}(2017{\natexlab{a}})\citenamefont {Sikorsky}, \citenamefont {Meir},
  \citenamefont {Ben-shlomi}, \citenamefont {Akerman},\ and\ \citenamefont
  {Ozeri}}]{Sikorsky2017a}%
  \BibitemOpen
  \bibfield  {author} {\bibinfo {author} {\bibfnamefont {Tomas}\ \bibnamefont
  {Sikorsky}}, \bibinfo {author} {\bibfnamefont {Ziv}\ \bibnamefont {Meir}},
  \bibinfo {author} {\bibfnamefont {Ruti}\ \bibnamefont {Ben-shlomi}}, \bibinfo
  {author} {\bibfnamefont {Nitzan}\ \bibnamefont {Akerman}}, \ and\ \bibinfo
  {author} {\bibfnamefont {Roee}\ \bibnamefont {Ozeri}},\ }\bibfield  {title}
  {\enquote {\bibinfo {title} {{Spin controlled atom-ion inelastic
  collisions}},}\ }\href {http://arxiv.org/abs/1709.00775} {\bibfield
  {journal} {\bibinfo  {journal} {ArXiv e-prints arXiv:1709.00775}\ } (\bibinfo
  {year} {2017}{\natexlab{a}})},\ \Eprint {http://arxiv.org/abs/1709.00775}
  {arXiv:1709.00775} \BibitemShut {NoStop}%
\bibitem [{\citenamefont {F{\"{u}}rst}\ \emph {et~al.}(2017)\citenamefont
  {F{\"{u}}rst}, \citenamefont {Feldker}, \citenamefont {Ewald}, \citenamefont
  {Joger}, \citenamefont {Tomza},\ and\ \citenamefont {Gerritsma}}]{Furst2017}%
  \BibitemOpen
  \bibfield  {author} {\bibinfo {author} {\bibfnamefont {Henning}\ \bibnamefont
  {F{\"{u}}rst}}, \bibinfo {author} {\bibfnamefont {Thomas}\ \bibnamefont
  {Feldker}}, \bibinfo {author} {\bibfnamefont {Norman~Vincenz}\ \bibnamefont
  {Ewald}}, \bibinfo {author} {\bibfnamefont {Jannis}\ \bibnamefont {Joger}},
  \bibinfo {author} {\bibfnamefont {Micha{\l}}\ \bibnamefont {Tomza}}, \ and\
  \bibinfo {author} {\bibfnamefont {Rene}\ \bibnamefont {Gerritsma}},\
  }\bibfield  {title} {\enquote {\bibinfo {title} {{Dynamics of a single ion
  spin impurity in a spin-polarized atomic bath}},}\ }\href
  {http://arxiv.org/abs/1712.07873} {\  (\bibinfo {year} {2017})},\ \Eprint
  {http://arxiv.org/abs/1712.07873} {arXiv:1712.07873} \BibitemShut {NoStop}%
\bibitem [{\citenamefont {Myatt}\ \emph {et~al.}(1997)\citenamefont {Myatt},
  \citenamefont {Burt}, \citenamefont {Ghrist}, \citenamefont {Cornell},\ and\
  \citenamefont {Wieman}}]{Myatt1997}%
  \BibitemOpen
  \bibfield  {author} {\bibinfo {author} {\bibfnamefont {C.~J.}\ \bibnamefont
  {Myatt}}, \bibinfo {author} {\bibfnamefont {E.~A.}\ \bibnamefont {Burt}},
  \bibinfo {author} {\bibfnamefont {R.~W.}\ \bibnamefont {Ghrist}}, \bibinfo
  {author} {\bibfnamefont {E.~A.}\ \bibnamefont {Cornell}}, \ and\ \bibinfo
  {author} {\bibfnamefont {C.~E.}\ \bibnamefont {Wieman}},\ }\bibfield  {title}
  {\enquote {\bibinfo {title} {{Production of two overlapping bose-einstein
  condensates by sympathetic cooling}},}\ }\href {\doibase
  10.1103/PhysRevLett.78.586} {\bibfield  {journal} {\bibinfo  {journal}
  {Physical Review Letters}\ }\textbf {\bibinfo {volume} {78}},\ \bibinfo
  {pages} {586--589} (\bibinfo {year} {1997})}\BibitemShut {NoStop}%
\bibitem [{\citenamefont {Truscott}(2001)}]{Truscott2001}%
  \BibitemOpen
  \bibfield  {author} {\bibinfo {author} {\bibfnamefont {A.~G.}\ \bibnamefont
  {Truscott}},\ }\bibfield  {title} {\enquote {\bibinfo {title} {{Observation
  of Fermi Pressure in a Gas of Trapped Atoms}},}\ }\href {\doibase
  10.1126/science.1059318} {\bibfield  {journal} {\bibinfo  {journal}
  {Science}\ }\textbf {\bibinfo {volume} {291}},\ \bibinfo {pages} {2570--2572}
  (\bibinfo {year} {2001})}\BibitemShut {NoStop}%
\bibitem [{\citenamefont {J{\"{o}}ckel}\ \emph {et~al.}(2015)\citenamefont
  {J{\"{o}}ckel}, \citenamefont {Faber}, \citenamefont {Kampschulte},
  \citenamefont {Korppi}, \citenamefont {Rakher},\ and\ \citenamefont
  {Treutlein}}]{Jockel2015}%
  \BibitemOpen
  \bibfield  {author} {\bibinfo {author} {\bibfnamefont {Andreas}\ \bibnamefont
  {J{\"{o}}ckel}}, \bibinfo {author} {\bibfnamefont {Aline}\ \bibnamefont
  {Faber}}, \bibinfo {author} {\bibfnamefont {Tobias}\ \bibnamefont
  {Kampschulte}}, \bibinfo {author} {\bibfnamefont {Maria}\ \bibnamefont
  {Korppi}}, \bibinfo {author} {\bibfnamefont {Matthew~T.}\ \bibnamefont
  {Rakher}}, \ and\ \bibinfo {author} {\bibfnamefont {Philipp}\ \bibnamefont
  {Treutlein}},\ }\bibfield  {title} {\enquote {\bibinfo {title} {{Sympathetic
  cooling of a membrane oscillator in a hybrid mechanical–atomic system}},}\
  }\href {\doibase 10.1038/nnano.2014.278} {\bibfield  {journal} {\bibinfo
  {journal} {Nature Nanotechnology}\ }\textbf {\bibinfo {volume} {10}},\
  \bibinfo {pages} {55--59} (\bibinfo {year} {2015})},\ \Eprint
  {http://arxiv.org/abs/1407.6820} {arXiv:1407.6820} \BibitemShut {NoStop}%
\bibitem [{\citenamefont {Zipkes}\ \emph {et~al.}(2010)\citenamefont {Zipkes},
  \citenamefont {Palzer}, \citenamefont {Sias},\ and\ \citenamefont
  {K{\"{o}}hl}}]{Zipkes2010}%
  \BibitemOpen
  \bibfield  {author} {\bibinfo {author} {\bibfnamefont {Christoph}\
  \bibnamefont {Zipkes}}, \bibinfo {author} {\bibfnamefont {Stefan}\
  \bibnamefont {Palzer}}, \bibinfo {author} {\bibfnamefont {Carlo}\
  \bibnamefont {Sias}}, \ and\ \bibinfo {author} {\bibfnamefont {Michael}\
  \bibnamefont {K{\"{o}}hl}},\ }\bibfield  {title} {\enquote {\bibinfo {title}
  {{A trapped single ion inside a Bose-Einstein condensate.}}}\ }\href
  {\doibase 10.1038/nature08865} {\bibfield  {journal} {\bibinfo  {journal}
  {Nature}\ }\textbf {\bibinfo {volume} {464}},\ \bibinfo {pages} {388--391}
  (\bibinfo {year} {2010})},\ \Eprint {http://arxiv.org/abs/1002.3304}
  {arXiv:1002.3304} \BibitemShut {NoStop}%
\bibitem [{\citenamefont {Schmid}\ \emph {et~al.}(2010)\citenamefont {Schmid},
  \citenamefont {H{\"{a}}rter},\ and\ \citenamefont {Denschlag}}]{Schmid2010}%
  \BibitemOpen
  \bibfield  {author} {\bibinfo {author} {\bibfnamefont {Stefan}\ \bibnamefont
  {Schmid}}, \bibinfo {author} {\bibfnamefont {Arne}\ \bibnamefont
  {H{\"{a}}rter}}, \ and\ \bibinfo {author} {\bibfnamefont {Johannes~Hecker}\
  \bibnamefont {Denschlag}},\ }\bibfield  {title} {\enquote {\bibinfo {title}
  {{Dynamics of a cold trapped ion in a Bose-Einstein condensate}},}\ }\href
  {\doibase 10.1103/PhysRevLett.105.133202} {\bibfield  {journal} {\bibinfo
  {journal} {Physical Review Letters}\ }\textbf {\bibinfo {volume} {105}},\
  \bibinfo {pages} {133202} (\bibinfo {year} {2010})},\ \Eprint
  {http://arxiv.org/abs/1007.4717} {arXiv:1007.4717} \BibitemShut {NoStop}%
\bibitem [{\citenamefont {Rellergert}\ \emph {et~al.}(2013)\citenamefont
  {Rellergert}, \citenamefont {Sullivan}, \citenamefont {Schowalter},
  \citenamefont {Kotochigova}, \citenamefont {Chen},\ and\ \citenamefont
  {Hudson}}]{Rellergert2013}%
  \BibitemOpen
  \bibfield  {author} {\bibinfo {author} {\bibfnamefont {Wade~G.}\ \bibnamefont
  {Rellergert}}, \bibinfo {author} {\bibfnamefont {Scott~T.}\ \bibnamefont
  {Sullivan}}, \bibinfo {author} {\bibfnamefont {Steven~J.}\ \bibnamefont
  {Schowalter}}, \bibinfo {author} {\bibfnamefont {Svetlana}\ \bibnamefont
  {Kotochigova}}, \bibinfo {author} {\bibfnamefont {Kuang}\ \bibnamefont
  {Chen}}, \ and\ \bibinfo {author} {\bibfnamefont {Eric~R.}\ \bibnamefont
  {Hudson}},\ }\bibfield  {title} {\enquote {\bibinfo {title} {{Evidence for
  sympathetic vibrational cooling of translationally cold molecules}},}\ }\href
  {\doibase 10.1038/nature11937} {\bibfield  {journal} {\bibinfo  {journal}
  {Nature}\ }\textbf {\bibinfo {volume} {495}},\ \bibinfo {pages} {490--494}
  (\bibinfo {year} {2013})}\BibitemShut {NoStop}%
\bibitem [{\citenamefont {Schowalter}\ \emph {et~al.}(2016)\citenamefont
  {Schowalter}, \citenamefont {Dunning}, \citenamefont {Chen}, \citenamefont
  {Puri}, \citenamefont {Schneider},\ and\ \citenamefont {Hudson}}]{BlueSky}%
  \BibitemOpen
  \bibfield  {author} {\bibinfo {author} {\bibfnamefont {Steven~J.}\
  \bibnamefont {Schowalter}}, \bibinfo {author} {\bibfnamefont {Alexander~J.}\
  \bibnamefont {Dunning}}, \bibinfo {author} {\bibfnamefont {Kuang}\
  \bibnamefont {Chen}}, \bibinfo {author} {\bibfnamefont {Prateek}\
  \bibnamefont {Puri}}, \bibinfo {author} {\bibfnamefont {Christian}\
  \bibnamefont {Schneider}}, \ and\ \bibinfo {author} {\bibfnamefont {Eric~R.}\
  \bibnamefont {Hudson}},\ }\bibfield  {title} {\enquote {\bibinfo {title}
  {{Blue-sky bifurcation of ion energies and the limits of neutral-gas
  sympathetic cooling of trapped ions}},}\ }\href {\doibase
  10.1038/ncomms12448} {\bibfield  {journal} {\bibinfo  {journal} {Nature
  Communications}\ }\textbf {\bibinfo {volume} {7}},\ \bibinfo {pages} {12448}
  (\bibinfo {year} {2016})},\ \Eprint {http://arxiv.org/abs/1608.03376}
  {arXiv:1608.03376} \BibitemShut {NoStop}%
\bibitem [{\citenamefont {Meir}\ \emph {et~al.}(2016)\citenamefont {Meir},
  \citenamefont {Sikorsky}, \citenamefont {Ben-shlomi}, \citenamefont
  {Akerman}, \citenamefont {Dallal},\ and\ \citenamefont {Ozeri}}]{Meir2016}%
  \BibitemOpen
  \bibfield  {author} {\bibinfo {author} {\bibfnamefont {Ziv}\ \bibnamefont
  {Meir}}, \bibinfo {author} {\bibfnamefont {Tomas}\ \bibnamefont {Sikorsky}},
  \bibinfo {author} {\bibfnamefont {Ruti}\ \bibnamefont {Ben-shlomi}}, \bibinfo
  {author} {\bibfnamefont {Nitzan}\ \bibnamefont {Akerman}}, \bibinfo {author}
  {\bibfnamefont {Yehonatan}\ \bibnamefont {Dallal}}, \ and\ \bibinfo {author}
  {\bibfnamefont {Roee}\ \bibnamefont {Ozeri}},\ }\bibfield  {title} {\enquote
  {\bibinfo {title} {{Dynamics of a Ground-State Cooled Ion Colliding with
  Ultracold Atoms}},}\ }\href {\doibase 10.1103/PhysRevLett.117.243401}
  {\bibfield  {journal} {\bibinfo  {journal} {Physical Review Letters}\
  }\textbf {\bibinfo {volume} {117}},\ \bibinfo {pages} {243401} (\bibinfo
  {year} {2016})}\BibitemShut {NoStop}%
\bibitem [{\citenamefont {Paul}(1990)}]{Paul1990}%
  \BibitemOpen
  \bibfield  {author} {\bibinfo {author} {\bibfnamefont {Wolfgang}\
  \bibnamefont {Paul}},\ }\bibfield  {title} {\enquote {\bibinfo {title}
  {{Electromagnetic traps for charged and neutral particles}},}\ }\href
  {\doibase 10.1103/RevModPhys.62.531} {\bibfield  {journal} {\bibinfo
  {journal} {Reviews of Modern Physics}\ }\textbf {\bibinfo {volume} {62}},\
  \bibinfo {pages} {531--540} (\bibinfo {year} {1990})}\BibitemShut {NoStop}%
\bibitem [{\citenamefont {Major}\ and\ \citenamefont
  {Dehmelt}(1968)}]{Dehmelt1968}%
  \BibitemOpen
  \bibfield  {author} {\bibinfo {author} {\bibfnamefont {F~G}\ \bibnamefont
  {Major}}\ and\ \bibinfo {author} {\bibfnamefont {H~G}\ \bibnamefont
  {Dehmelt}},\ }\bibfield  {title} {\enquote {\bibinfo {title}
  {{Exchange-Collision Technique for the rf Spectroscopy of Stored Ions}},}\
  }\href {\doibase 10.1103/PhysRev.170.91} {\bibfield  {journal} {\bibinfo
  {journal} {Physical Review}\ }\textbf {\bibinfo {volume} {170}},\ \bibinfo
  {pages} {91--107} (\bibinfo {year} {1968})}\BibitemShut {NoStop}%
\bibitem [{\citenamefont {DeVoe}(2009)}]{DeVoe2009}%
  \BibitemOpen
  \bibfield  {author} {\bibinfo {author} {\bibfnamefont {Ralph~G.}\
  \bibnamefont {DeVoe}},\ }\bibfield  {title} {\enquote {\bibinfo {title}
  {{Power-Law Distributions for a Trapped Ion Interacting with a Classical
  Buffer Gas}},}\ }\href {\doibase 10.1103/PhysRevLett.102.063001} {\bibfield
  {journal} {\bibinfo  {journal} {Physical Review Letters}\ }\textbf {\bibinfo
  {volume} {102}},\ \bibinfo {pages} {063001} (\bibinfo {year} {2009})},\
  \Eprint {http://arxiv.org/abs/0903.0637} {arXiv:0903.0637} \BibitemShut
  {NoStop}%
\bibitem [{\citenamefont {Zipkes}\ \emph {et~al.}(2011)\citenamefont {Zipkes},
  \citenamefont {Ratschbacher}, \citenamefont {Sias},\ and\ \citenamefont
  {K{\"{o}}hl}}]{Zipkes2011}%
  \BibitemOpen
  \bibfield  {author} {\bibinfo {author} {\bibfnamefont {Christoph}\
  \bibnamefont {Zipkes}}, \bibinfo {author} {\bibfnamefont {Lothar}\
  \bibnamefont {Ratschbacher}}, \bibinfo {author} {\bibfnamefont {Carlo}\
  \bibnamefont {Sias}}, \ and\ \bibinfo {author} {\bibfnamefont {Michael}\
  \bibnamefont {K{\"{o}}hl}},\ }\bibfield  {title} {\enquote {\bibinfo {title}
  {{Kinetics of a single trapped ion in an ultracold buffer gas}},}\
  }\href@noop {} {\bibfield  {journal} {\bibinfo  {journal} {New Journal of
  Physics}\ }\textbf {\bibinfo {volume} {13}},\ \bibinfo {pages} {53020}
  (\bibinfo {year} {2011})}\BibitemShut {NoStop}%
\bibitem [{\citenamefont {Chen}\ \emph {et~al.}(2014)\citenamefont {Chen},
  \citenamefont {Sullivan},\ and\ \citenamefont {Hudson}}]{Chen2014}%
  \BibitemOpen
  \bibfield  {author} {\bibinfo {author} {\bibfnamefont {Kuang}\ \bibnamefont
  {Chen}}, \bibinfo {author} {\bibfnamefont {Scott~T.}\ \bibnamefont
  {Sullivan}}, \ and\ \bibinfo {author} {\bibfnamefont {Eric~R.}\ \bibnamefont
  {Hudson}},\ }\bibfield  {title} {\enquote {\bibinfo {title} {{Neutral Gas
  Sympathetic Cooling of an Ion in a Paul Trap}},}\ }\href {\doibase
  10.1103/PhysRevLett.112.143009} {\bibfield  {journal} {\bibinfo  {journal}
  {Physical Review Letters}\ }\textbf {\bibinfo {volume} {112}},\ \bibinfo
  {pages} {143009} (\bibinfo {year} {2014})},\ \Eprint
  {http://arxiv.org/abs/arXiv:1310.5190v1} {arXiv:arXiv:1310.5190v1}
  \BibitemShut {NoStop}%
\bibitem [{\citenamefont {H{\"{o}}ltkemeier}\ \emph {et~al.}(2016)\citenamefont
  {H{\"{o}}ltkemeier}, \citenamefont {Weckesser}, \citenamefont
  {L{\'{o}}pez-Carrera},\ and\ \citenamefont {Weidem{\"{u}}ller}}]{Pascal2016}%
  \BibitemOpen
  \bibfield  {author} {\bibinfo {author} {\bibfnamefont {Bastian}\ \bibnamefont
  {H{\"{o}}ltkemeier}}, \bibinfo {author} {\bibfnamefont {Pascal}\ \bibnamefont
  {Weckesser}}, \bibinfo {author} {\bibfnamefont {Henry}\ \bibnamefont
  {L{\'{o}}pez-Carrera}}, \ and\ \bibinfo {author} {\bibfnamefont {Matthias}\
  \bibnamefont {Weidem{\"{u}}ller}},\ }\bibfield  {title} {\enquote {\bibinfo
  {title} {{Buffer-Gas Cooling of a Single Ion in a Multipole Radio Frequency
  Trap Beyond the Critical Mass Ratio}},}\ }\href {\doibase
  10.1103/PhysRevLett.116.233003} {\bibfield  {journal} {\bibinfo  {journal}
  {Physical Review Letters}\ }\textbf {\bibinfo {volume} {116}},\ \bibinfo
  {pages} {233003} (\bibinfo {year} {2016})},\ \Eprint
  {http://arxiv.org/abs/1505.06909} {arXiv:1505.06909} \BibitemShut {NoStop}%
\bibitem [{\citenamefont {Rouse}\ and\ \citenamefont
  {Willitsch}(2017)}]{Rouse2017}%
  \BibitemOpen
  \bibfield  {author} {\bibinfo {author} {\bibfnamefont {I.}~\bibnamefont
  {Rouse}}\ and\ \bibinfo {author} {\bibfnamefont {S.}~\bibnamefont
  {Willitsch}},\ }\bibfield  {title} {\enquote {\bibinfo {title}
  {{Superstatistical Energy Distributions of an Ion in an Ultracold Buffer
  Gas}},}\ }\href {\doibase 10.1103/PhysRevLett.118.143401} {\bibfield
  {journal} {\bibinfo  {journal} {Physical Review Letters}\ }\textbf {\bibinfo
  {volume} {118}},\ \bibinfo {pages} {143401} (\bibinfo {year} {2017})},\
  \Eprint {http://arxiv.org/abs/1703.06006} {arXiv:1703.06006} \BibitemShut
  {NoStop}%
\bibitem [{\citenamefont {Meir}\ \emph {et~al.}(2018)\citenamefont {Meir},
  \citenamefont {Sikorsky}, \citenamefont {Ben-shlomi}, \citenamefont
  {Akerman}, \citenamefont {Pinkas}, \citenamefont {Dallal},\ and\
  \citenamefont {Ozeri}}]{Meir2017exp}%
  \BibitemOpen
  \bibfield  {author} {\bibinfo {author} {\bibfnamefont {Ziv}\ \bibnamefont
  {Meir}}, \bibinfo {author} {\bibfnamefont {Tomas}\ \bibnamefont {Sikorsky}},
  \bibinfo {author} {\bibfnamefont {Ruti}\ \bibnamefont {Ben-shlomi}}, \bibinfo
  {author} {\bibfnamefont {Nitzan}\ \bibnamefont {Akerman}}, \bibinfo {author}
  {\bibfnamefont {Meirav}\ \bibnamefont {Pinkas}}, \bibinfo {author}
  {\bibfnamefont {Yehonatan}\ \bibnamefont {Dallal}}, \ and\ \bibinfo {author}
  {\bibfnamefont {Roee}\ \bibnamefont {Ozeri}},\ }\bibfield  {title} {\enquote
  {\bibinfo {title} {{Experimental apparatus for overlapping a ground-state
  cooled ion with ultracold atoms}},}\ }\href {\doibase
  10.1080/09500340.2017.1397217} {\bibfield  {journal} {\bibinfo  {journal}
  {Journal of Modern Optics}\ }\textbf {\bibinfo {volume} {65}},\ \bibinfo
  {pages} {387--405} (\bibinfo {year} {2018})}\BibitemShut {NoStop}%
\bibitem [{\citenamefont {Meir}\ \emph {et~al.}(2017)\citenamefont {Meir},
  \citenamefont {Sikorsky}, \citenamefont {Akerman}, \citenamefont
  {Ben-shlomi}, \citenamefont {Pinkas},\ and\ \citenamefont
  {Ozeri}}]{Meir2017cooling}%
  \BibitemOpen
  \bibfield  {author} {\bibinfo {author} {\bibfnamefont {Ziv}\ \bibnamefont
  {Meir}}, \bibinfo {author} {\bibfnamefont {Tomas}\ \bibnamefont {Sikorsky}},
  \bibinfo {author} {\bibfnamefont {Nitzan}\ \bibnamefont {Akerman}}, \bibinfo
  {author} {\bibfnamefont {Ruti}\ \bibnamefont {Ben-shlomi}}, \bibinfo {author}
  {\bibfnamefont {Meirav}\ \bibnamefont {Pinkas}}, \ and\ \bibinfo {author}
  {\bibfnamefont {Roee}\ \bibnamefont {Ozeri}},\ }\bibfield  {title} {\enquote
  {\bibinfo {title} {{Single-shot energy measurement of a single atom and the
  direct reconstruction of its energy distribution}},}\ }\href {\doibase
  10.1103/PhysRevA.96.020701} {\bibfield  {journal} {\bibinfo  {journal}
  {Physical Review A}\ }\textbf {\bibinfo {volume} {96}},\ \bibinfo {pages}
  {020701} (\bibinfo {year} {2017})},\ \Eprint
  {http://arxiv.org/abs/1706.00858} {arXiv:1706.00858} \BibitemShut {NoStop}%
\bibitem [{\citenamefont {Wesenberg}\ \emph {et~al.}(2007)\citenamefont
  {Wesenberg}, \citenamefont {Epstein}, \citenamefont {Leibfried},
  \citenamefont {Blakestad}, \citenamefont {Britton}, \citenamefont {Home},
  \citenamefont {Itano}, \citenamefont {Jost}, \citenamefont {Knill},
  \citenamefont {Langer}, \citenamefont {Ozeri}, \citenamefont {Seidelin},\
  and\ \citenamefont {Wineland}}]{Wesenberg2007}%
  \BibitemOpen
  \bibfield  {author} {\bibinfo {author} {\bibfnamefont {J.~H.}\ \bibnamefont
  {Wesenberg}}, \bibinfo {author} {\bibfnamefont {R.~J.}\ \bibnamefont
  {Epstein}}, \bibinfo {author} {\bibfnamefont {D.}~\bibnamefont {Leibfried}},
  \bibinfo {author} {\bibfnamefont {R.~B.}\ \bibnamefont {Blakestad}}, \bibinfo
  {author} {\bibfnamefont {J.}~\bibnamefont {Britton}}, \bibinfo {author}
  {\bibfnamefont {J.~P.}\ \bibnamefont {Home}}, \bibinfo {author}
  {\bibfnamefont {W.~M.}\ \bibnamefont {Itano}}, \bibinfo {author}
  {\bibfnamefont {J.~D.}\ \bibnamefont {Jost}}, \bibinfo {author}
  {\bibfnamefont {E.}~\bibnamefont {Knill}}, \bibinfo {author} {\bibfnamefont
  {C.}~\bibnamefont {Langer}}, \bibinfo {author} {\bibfnamefont
  {R.}~\bibnamefont {Ozeri}}, \bibinfo {author} {\bibfnamefont
  {S.}~\bibnamefont {Seidelin}}, \ and\ \bibinfo {author} {\bibfnamefont
  {D.~J.}\ \bibnamefont {Wineland}},\ }\bibfield  {title} {\enquote {\bibinfo
  {title} {{Fluorescence during Doppler cooling of a single trapped atom}},}\
  }\href {\doibase 10.1103/PhysRevA.76.053416} {\bibfield  {journal} {\bibinfo
  {journal} {Physical Review A}\ }\textbf {\bibinfo {volume} {76}},\ \bibinfo
  {pages} {053416} (\bibinfo {year} {2007})},\ \Eprint
  {http://arxiv.org/abs/0707.1314} {arXiv:0707.1314} \BibitemShut {NoStop}%
\bibitem [{\citenamefont {Sikorsky}\ \emph
  {et~al.}(2017{\natexlab{b}})\citenamefont {Sikorsky}, \citenamefont {Meir},
  \citenamefont {Akerman}, \citenamefont {Ben-shlomi},\ and\ \citenamefont
  {Ozeri}}]{Sikorsky2017}%
  \BibitemOpen
  \bibfield  {author} {\bibinfo {author} {\bibfnamefont {Tomas}\ \bibnamefont
  {Sikorsky}}, \bibinfo {author} {\bibfnamefont {Ziv}\ \bibnamefont {Meir}},
  \bibinfo {author} {\bibfnamefont {Nitzan}\ \bibnamefont {Akerman}}, \bibinfo
  {author} {\bibfnamefont {Ruti}\ \bibnamefont {Ben-shlomi}}, \ and\ \bibinfo
  {author} {\bibfnamefont {Roee}\ \bibnamefont {Ozeri}},\ }\bibfield  {title}
  {\enquote {\bibinfo {title} {{Doppler cooling thermometry of a multilevel ion
  in the presence of micromotion}},}\ }\href {\doibase
  10.1103/PhysRevA.96.012519} {\bibfield  {journal} {\bibinfo  {journal}
  {Physical Review A}\ }\textbf {\bibinfo {volume} {96}},\ \bibinfo {pages}
  {012519} (\bibinfo {year} {2017}{\natexlab{b}})},\ \Eprint
  {http://arxiv.org/abs/1705.00453} {arXiv:1705.00453} \BibitemShut {NoStop}%
\bibitem [{\citenamefont {Dutta}\ \emph {et~al.}(2017)\citenamefont {Dutta},
  \citenamefont {Sawant},\ and\ \citenamefont {Rangwala}}]{Dutta2017}%
  \BibitemOpen
  \bibfield  {author} {\bibinfo {author} {\bibfnamefont {Sourav}\ \bibnamefont
  {Dutta}}, \bibinfo {author} {\bibfnamefont {Rahul}\ \bibnamefont {Sawant}}, \
  and\ \bibinfo {author} {\bibfnamefont {S.~A.}\ \bibnamefont {Rangwala}},\
  }\bibfield  {title} {\enquote {\bibinfo {title} {{Collisional Cooling of
  Light Ions by Cotrapped Heavy Atoms}},}\ }\href {\doibase
  10.1103/PhysRevLett.118.113401} {\bibfield  {journal} {\bibinfo  {journal}
  {Physical Review Letters}\ }\textbf {\bibinfo {volume} {118}},\ \bibinfo
  {pages} {113401} (\bibinfo {year} {2017})}\BibitemShut {NoStop}%
\bibitem [{SM()}]{SM}%
  \BibitemOpen
  \bibfield  {title} {\enquote {\bibinfo {title} {{See Supplemental Material
  online for details regarding experimental details, Likelihood ratio test
  stochastic simulation and scattering angle calculation.}}}\ }\href@noop {} {\
  }\BibitemShut {NoStop}%
\bibitem [{\citenamefont {Cirac}\ \emph {et~al.}(1994)\citenamefont {Cirac},
  \citenamefont {Garay}, \citenamefont {Blatt}, \citenamefont {Parkins},\ and\
  \citenamefont {Zoller}}]{Cirac1994}%
  \BibitemOpen
  \bibfield  {author} {\bibinfo {author} {\bibfnamefont {J.~I.}\ \bibnamefont
  {Cirac}}, \bibinfo {author} {\bibfnamefont {L.~J.}\ \bibnamefont {Garay}},
  \bibinfo {author} {\bibfnamefont {R.}~\bibnamefont {Blatt}}, \bibinfo
  {author} {\bibfnamefont {A.~S.}\ \bibnamefont {Parkins}}, \ and\ \bibinfo
  {author} {\bibfnamefont {P.}~\bibnamefont {Zoller}},\ }\bibfield  {title}
  {\enquote {\bibinfo {title} {{Laser cooling of trapped ions: The influence of
  micromotion}},}\ }\href {\doibase 10.1103/PhysRevA.49.421} {\bibfield
  {journal} {\bibinfo  {journal} {Physical Review A}\ }\textbf {\bibinfo
  {volume} {49}},\ \bibinfo {pages} {421--432} (\bibinfo {year}
  {1994})}\BibitemShut {NoStop}%
\end{thebibliography}%

\section{Supplemental material}
\subsection{Interaction time calculation}
We measured the atomic profile using time-of-flight absorption imaging. The atomic cloud is cigar-shaped with approximately Gaussian density, $n_{at}=n_0e^{-x^2/2\sigma_y^2-y^2/2\sigma_z^2-z^2/2\sigma_x^2}$ distribution in both radial, $\sigma_x=\sigma_y$=5 $\mu$m, and axial $\sigma_{z}$=70 $\mu$m axes. We measured the peak atomic density to be, $n_0=4\cdot10^{11}$ cm$^{-3}$ and the number of atoms in our trap is approximately 7,500.

The Langevin rate of collisions, $\Gamma=2 \pi n_{at} \sqrt{C_4/\mu}$, is independent of the collision energy. For our atomic experimental parameters of, $C_4=1.1\cdot10^{-56}$ J$\cdot$m$^4$ and $\mu\approx43.7$ amu, it yields a collision rate, $\Gamma$=0.9 kHz, for the atomic peak density.

In the first axial experiment, we prepare the atoms 50 $\mu$m vertically above the center of the ion trap. We then transport the atoms to overlap with the center of the trap using a piezo driven mirror which controls the position of the horizontal atoms' crossed dipole trap beam. The atoms' velocity during the transport is approximately constant, $v_{t}$=0.5 $\mu$m/ms. It takes the atoms 100 ms to reach the center of the trap. 

Due to the transport, the ion interacts first with the outskirts of the atomic profile which leads to much reduced and time dependent collision rate. To take into account the effect of the transport, we calculate the effective interaction time, $t_\textrm{eff}$, as if the atoms overlap the center of the trap from the beginning of the experiment,
\begin{equation}
	t_\textrm{eff}=\int\limits_{-\infty}^{t_t-x_t/v_t}e^{-\frac{(v_{t}t)^2}{2\sigma_y^2}}dt=t_0\left( \textrm{erf}\left( \frac{v_t t_t - x_t}{\sqrt{2}\sigma_y}\right) + 1 \right).
\end{equation}
Here, $0 \le t_t \le 100$ ms is the transport time, $x_t$=50 $\mu$m is the transport distance and $t_0=\sqrt{\frac{\pi}{2}}\frac{\sigma_y}{v_t}=9.2$ ms is the effective time for full transport, $t_t$=100 ms. The experimental interaction time, given in Fig. 2a-f, is then calculated by,
\begin{equation}
	t_\textrm{int.}=
	    \begin{cases}
			\begin{aligned}
				&t_\textrm{eff} & ,0 \le t_t \le x_t/v_t\\
				&t_0+(t_t-x_t/v_t) & ,t_t \ge x_t/v_t.
    		\end{aligned}
		\end{cases}
\end{equation}
For experiments with only single to few collisions (Fig. 2b-e in the main text) we stop the transport and release the atoms before they reach the center of the trap.

Our transport calculation takes into account the motion of the atoms with respect to the center of the ion trap. However, the ion is prepared in a classical coherent state and is preforming an oscillatory motion with an amplitude of $z_\textrm{ion}=71$ $\mu$m in the axial direction relative to the trap center. This motion amplitude is comparable to the atomic cloud size. We numerically calculate the effective reduced density due to this motion, $n_\textrm{eff}=0.79\cdot n_0$, which yields a slower Langevin collision rate, $\Gamma_\textrm{eff}$=0.71 kHz.

In the second, radial, experiment, we compare the experimental time to simulation time directly by implementing the motion of the atomic cloud into the simulation. 

\subsection{Calculation of the energy in the axial and radial modes of the trap}
\begin{figure}
	\centering
	\includegraphics[width=\linewidth,trim={0cm 0cm 1cm 0cm},clip]{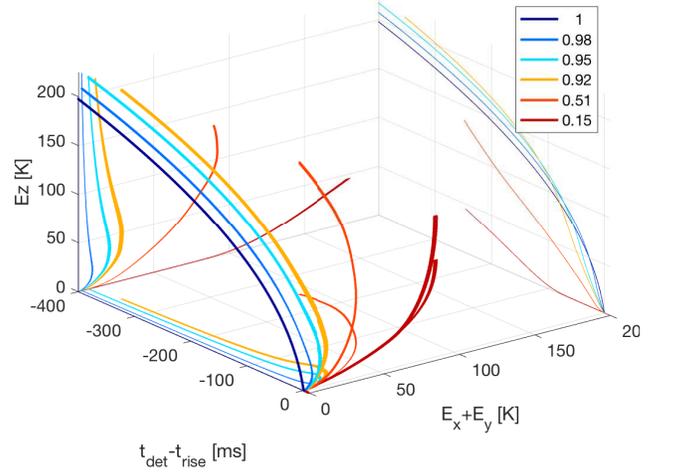}\\
	\caption{\textbf{Numerical calculation of the ion's energy dynamics during cooling}. The energy of the ion in the axial (z-axis) and radial (y-axis) is plotted as function of the time for reaching the Doppler cooling temperature (x-axis). Color represent different axial to radial energy ratios (legend).}
	\label{fig:energy3d}
\end{figure}

The results of a numerical calculation for initial energy of 225K with different initial axial to radial energy distribution is given in Fig. \ref{fig:energy3d}. This is the same numerical calculation which was used throughout the main text. Here, we present the energy dynamics separated to the axial (z) and radial (x+y) modes of the trap. 

\subsection{Likelihood ratio test}
We calculate the likelihood, $\mathcal{L}_i^m$, of a single data instance, $x^m(t)$, to be described by each of the numerical trajectories, $\theta_i$:
\begin{equation}
    \mathcal{L}_i^m\left(x^m(t)|\theta_i\right)=\prod_t \frac{\theta_i(t)^{x^m(t)}e^{-\theta_i(t)}}{x^m(t)!}.
\end{equation}
Here, the trajectory, $\theta(t)$, is sampled at the experimental sampling times due to the 1 ms binning. The subscript, $i$, indicates which numerical calculation is used to calculate the likelihood. The superscript $m$ indicates what instance of data is evaluated. In the LRT we compare the likelihoods values from all the numerical calculations, $i$, for each experimental data, $m$, and pick the one with the maximal value: 
\begin{equation}
    \mathcal{L}_i^m\left(x^m|\theta_i\right)>\mathcal{L}_j^m\left(x^m|\theta_j\right) \forall j\neq i.
\end{equation}
We performed a stochastic numerical simulation to analyze the LRT method (see details in the following paragraph). The type I errors for wrongly detecting mode $i$ as mode $j$ are less than 1\% when the cooling trajectory is longer than 100 ms. We use the numerical calculation, $i$, chosen by the LRT to extract the energy of the ion for the experimental instance, $m$.

We used the results of the numerical calculation for different initial energy distributions between the axial and radial modes of the trap (Fig. 1b in the main text) to test our LRT analysis. We sampled the numerically calculated fluorescence trajectories according to the experimental photon-counting binning of 1ms and created a simulated stochastic data using a Poisson random distribution. We then performed the LRT analysis on the simulated data (100,000 trajectories for each numerical calculation) and determined the type I error due to Poisson noise.

\begin{figure}
	\centering
	\includegraphics[width=\linewidth,trim={0cm 0cm 1cm 0cm},clip]{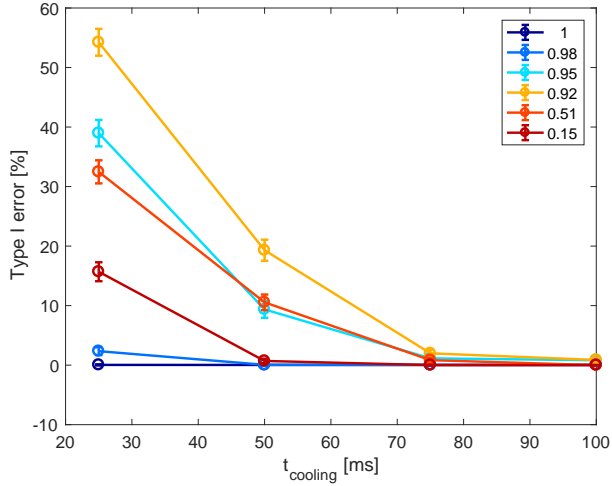}\\
	\caption{\textbf{Stochastic simulation of the LRT.} Type I error as function of the data span used for the LRT. Different colors indicates from which numerical simulation of different axial to radial energy ratio (legend) the stochastic data was created.}
	\label{fig:likelihoodspan}
\end{figure}

When performing likelihood analysis to single-shot data we have the freedom to choose the time span on which we perform the likelihood. We usually choose a span from -100 ms to 10 ms which captures all the discrepancies of the different numerical calculations (see Fig. 1b of the main text). However, 100 ms cooling time corresponds to initial ion's energy of $\sim$100K. In order to perform LRT on colder events we need to reduce the likelihood span. In Fig. \ref{fig:likelihoodspan} we show the results of the stochastic simulation for different likelihood spans. We see that as the likelihood span reduces, the percentage of type I errors increases. For that, we choose a minimal span of 50 ms cooling time which limits our LRT analysis to $\sim$50 K. Below this energy we estimate the energy using the the numerical simulation with the initial energy ratio of Ez/Er=0.15.

\subsection{Calculation of the scattering angle distribution}
We consider an ion with mass, m, and a velocity, v, which we choose to coincide with the z-axis of the trap, and an atom with mass, M, at rest in the lab-frame. For simplicity, we first consider the case in which m=M. In the center-of-mass frame, the ion's velocity is v/2 and the atom's velocity is -v/2 along the z-axis. After an elastic collision which preserves both momentum and energy, the ion's and atom's velocities remain v/2 with opposite directions, however, at angle $\theta$ from the z-axis. Since in head-on collision there is an equal probability to scatter into 4$\pi$, $\cos(\theta)$ is uniformly distributed between -1 to 1 where we implicitly choose the scattering angle $\phi$ to be zero.

In the lab frame, the velocity of the ion after a collision is given by:
\begin{equation}
    v/2(1+\cos(\theta))\hat{z}+v/2\sin(\theta)\hat{x}.
\end{equation}
Here, we rotate the radial axes such that the scattering products lie along the x-axis. The total energy of the ion before a collision, $E_b$, is:
\begin{equation}
    E_b=mv^2/2+mz^2\omega_z^2/2,
\end{equation}
where $z$ is the (unknown) position of the ion in the harmonic z-axis of the Paul trap and $\omega_z$ is the trap's harmonic frequency. The ion's energy after a collision, $E_a$, is given by:
\begin{equation}
    E_a=m(v/2)^2(1+\cos(\theta))^2/2+m(v/2)^2\sin^2(\theta)/2+mz^2\omega_z^2.
\end{equation}
Here, we assume that the collision is instantaneous such that the position of the collision, (0,0,z), is not changed during the collision. For that, there is no contribution from the radial trapping potential. To get rid of the dependence of the energy on the collision position, z, we look on the energy difference during a collision:
\begin{equation}\label{eq:Ediff}
    \Delta E=E_b-E_a=m(v/2)^2(1-\cos(\theta)).
\end{equation}
This expression depends on the ion's initial velocity, v, which is also unknown since the ion is in an harmonic potential (only the initial energy is known). To get rid of this dependence we divide the energy difference with the radial energy:
\begin{equation}
    \frac{\Delta E}{E_r}=2\frac{1-\cos(\theta)}{1-\cos^2(\theta)}.
\end{equation}
For the general case of an ion with mass m and an atom with mass M, this expression only slightly changes:
\begin{equation}\label{eq:1}
    \frac{\Delta E}{E_r}=2\frac{m}{M}\frac{1-\cos(\theta)}{1-\cos^2(\theta)}.
\end{equation}
We can solve this second order polynomial equation to get an expression for $\cos(\theta)$ which is given in Eq. \ref{eq:costheta} of the main text.

\begin{figure}
	\centering
	\includegraphics[width=\linewidth,trim={0cm 0cm 1cm 0cm},clip]{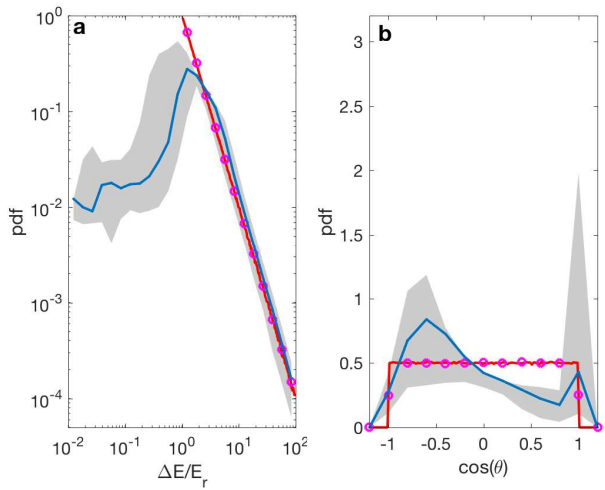}\\
	\caption{\textbf{Scattering angle simulation.} Histograms of the energy difference divided by the radial energy after a collision (a) and the cosine of the scattering angle (b). Red lines are the theoretical prediction of Langevin scattering. Magenta circles are MD simulation results of exactly single collision of ion initialized in the axial direction with zero temperature atoms and no excess micromotion. The blue line is the same simulation, however, in the analysis we take into account the experimental limitations in the data analysis described in the text. Gray areas gives the uncertainty region for the data analysis.}
	\label{fig:scattering_angle_sm}
\end{figure}

The red line in Fig. 3 of the main text is the Langevin scattering angle distribution given in terms of Eq. \ref{eq:1} (Fig. 3a) and $\cos(\theta)$ (Fig. 3b). We can reproduce this theoretical prediction exactly using our MD simulation as can be seen in Fig. \ref{fig:scattering_angle_sm} by the magenta circles. However, since each experimental measurement of the energy is destructive, we cannot obtain the exact value of the energy difference in a collision given in Eq. \ref{eq:Ediff}. Instead, we use the mean value of the initial energy, $\left<E_b\right>=183.7$ K, to get an approximated value of the energy difference,
\begin{equation}\label{eq:Ediffmean}
    \left<\Delta E\right>=\left<E_b\right>-E_a.
\end{equation}
The error for $\left<\Delta E\right>$ is given by the standard deviation of the initial energy, $\Delta \left<E_b\right>=2.3$ K. 

Another error source is due to the binning of the radial energy due to the LRT analysis. We bin the radial energy to the same binning used in the experimental data analysis. We calculate $E_z/E_a$ of each simulation run. For $E_z/E_a>0.9923$ we consider the event as no collision ($E_z/E_a=1)$. For $E_z/E_a$ values in the ranges given by the vector,  (0.9923, 0.9690, 0.9375, 0.7134, 0.3284, 0), we bin $Ez/E_a$=(0.9846, 0.9534, 0.9216, 0.5053, 0.1515). The radial energy is given by, $E_r=E_a(1-E_z/E_a)$. The results of this binning together with the mean energy effect is given by the blue line in Fig. \ref{fig:scattering_angle_sm} and Fig. 3 of the main text. The gray area indicates the values the simulation can vary when taking the edges instead of the mean value of each bin.

\end{document}